\documentclass[a4paper,conference]{IEEEtran}
\usepackage[latin9]{inputenc}
\usepackage{float}
\usepackage{units}
\usepackage{amstext}
\usepackage{graphicx}
\PassOptionsToPackage{normalem}{ulem}
\usepackage{ulem}

\makeatletter

\pdfpageheight\paperheight
\pdfpagewidth\paperwidth

\providecommand{\tabularnewline}{\\}
\floatstyle{ruled}
\newfloat{algorithm}{tbp}{loa}
\providecommand{\algorithmname}{Algorithm}
\floatname{algorithm}{\protect\algorithmname}

\usepackage{algorithmic}
\usepackage{amssymb}

\makeatother

\begin{document}

\title{BitSurfing: Wireless Communications with Outsourced Symbol Generation}

\author{Ageliki Tsioliaridou, Christos Liaskos, Sotiris Ioannidis\\
FORTH, Greece, \{atsiolia, cliaskos, sotiris\}@ics.forth.gr \thanks{This work was funded by the European Union via the Horizon 2020: Future
Emerging Topics call (FETOPEN), grant EU736876, project VISORSURF
(http://www.visorsurf.eu).}\\
\vspace{-20bp}
}
\maketitle
\begin{abstract}
Nano-IoT enables a wide range of ground-breaking technologies, but
face implementation challenges due to the extremity of the scale.
Space restrictions pose severe power supply considerations, to the
point where just a few packet transmissions are sufficient to deplete
state-of-the-art supplies. In turn, this translates to difficulties
in developing efficient protocols even for basic operations, such
as addressing and routing. The present work proposes a new network
adapter architecture that can address these challenges. The BitSurfing
adapter does not generate packets and, hence, abolishes the need for
the corresponding transmission circuitry and power consumption. Instead,
it relies on an external symbol generator. The BitSurfing adapter
reads incoming symbols, waiting for intended messages to appear in
the symbol stream. A short (1-bit), low-energy pulse is then emitted
to notify neighboring nodes. BitSurfing adapters are shown to exhibit
perpetual (and even battery-less) operation, ability to operate without
medium access control, while being completely transparent to applications.
Moreover, their operation is event-driven, allowing for clock-less
implementations. The new adapters are evaluated in a simulated multi-hop
nano-IoT network and are shown to offer nearly-perfect packet delivery
rates and practically no collisions, under any congestion level.
\end{abstract}


\section{Introduction\label{sec:Introduction}}

Nano-IoT will extend the reach of smart control to the level of molecules
and cells, with unprecedented impact in medicine and material manufacturing.
Combating diseases within the human body via autonomous nanomachines,
self-healing and self-monitoring materials are a few of the most visionary
applications. Materials with software-defined electromagnetic behavior
constitute applications under development, paving the way for programmable
wireless environments~\cite{cacm}.

Nanonode architectures under research, especially the ones targeting
the most visionary applications, seek to implement common components
(receiver, transmitter, processor, memory, battery, sensor and actuator)
down at the nanoscale. Among them, transceivers and batteries constitute
the major roadblocks for manufacturing nanonodes~\cite{Jornet.2012b}.
Regarding the transceiver, studies have identified the $THz$ band
and graphene antennas as the most promising approaches~\cite{Jornet.2012b}.
Nonetheless, their energy consumption constitutes a major problem.
Even exotic (hard to integrate) power supplies relying on energy harvesting
can only scavenge energy for $1$ packet transmission per approximately
$10$ sec~\cite{Jornet.2012b}. This makes the development of even
basic protocols\textendash such as addressing and routing\textendash highly
challenging~\cite{limitedSpace}.

The present work contributes a novel network adapter for nano-IoT\textendash the
BitSurfer\textendash that can overcome these shortcomings. Its operating
principle is unconventional, as it does not include creating and transmitting
packets and, hence, no corresponding circuitry. Instead, the task
of symbol creation is \emph{outsourced} to an external entity, the
symbol source. BitSurfing adapters simply listen to the symbol stream
created by the source, and await for desired messages to appear within
it. Simple 1-bit, low energy pulses are then used for notifying neighboring
nodes to read their streams and extract the message intended for them.
The present study details the benefits of BitSurfing adapters, which
include transparency, perpetual, battery-less (by prospect) and Medium
Access Control (MAC)-less operation. Moreover, the BitSurfing workflow
is completely event-driven, meaning that they can be implemented as
asynchronous chips~\cite{beerel2011proteus}. Since power is not
drawn on synchronized clock edges, the peak-power draw is much lower
and better suited for battery-less systems. Furthermore this can favor
miniaturization further, since common clock components\textendash such
as crystals and oscillators\textendash are sizable, energy consuming
(i.e., per tick) and one of the least-studied roadblocks towards manufacturing
nano-IoT networks. BitSurfing is enabled by recent hardware advances,
which have shown that wireless reception and data processing can be
perpetually powered by the received carrier itself, without the need
for a battery or any other power supply~\cite{Tabesh.}.

The paper is organized as follows. Section~\ref{sec:Background}
provides the prerequisites. Section~\ref{sec:System-Model} details
the BitSurfing architecture, while Section~\ref{sec:systemdesign}
presents its design workflow. Evaluation takes place in Section~\ref{sec:sim}.
Discussion and research directions are given in Section~\ref{sec:directions}.
The conclusion follows in Section~\ref{sec:Conclusion}.

\section{Preliminaries\label{sec:Background}}

A core element of the BitSurfing model is the notion of \emph{word
cover time}, i.e., the expected time to find a specific string in
an i.i.d. random binary sequence (\emph{bit stream}). This problem
has re-surfaced several times in the literature~\cite{McConnel.2001},
leading to the following outcomes.

Let $w$ denote a \emph{word}, i.e., a specific binary sequence of
'1's and '0's. Moreover, let $\left\Vert w\right\Vert $ denote the
size of the word $w$ measured in number of bits. Then, the cover
time $CT\left(w\right)$ of word $w$ is~\cite{McConnel.2001}:
\begin{equation}
CT\left(w\right)=2^{\left\Vert w\right\Vert }+2^{\left\Vert f\left(w\right)\right\Vert }+2^{\left\Vert f\left(f\left(w\right)\right)\right\Vert }+\ldots\label{eq:CTw}
\end{equation}
where $f\left(w\right)$ is the failure function, defined as the longest
prefix of $w$ that is also its suffix. (By definition, $f\left(w\right)$
returns an empty string if no such prefix exists). The summation in
equation (\ref{eq:CTw}) continues as long as the exponent is greater
than zero. A C-language implementation of equation (\ref{eq:CTw})
can be found in the literature~\cite{McConnel.1994}.

When we are not interested in a specific word, but rather for the
cover time of any word of size $\left\Vert w\right\Vert $ on average,
the expression is simplified to $CT\left(\left\Vert w\right\Vert \right)\sim\sqrt{\pi\cdot2^{\left\Vert w\right\Vert }},$
when $\left\Vert w\right\Vert \to\infty$~\cite{McConnel.2001}.
For small $\left\Vert w\right\Vert $ values (e.g., $\left\Vert w\right\Vert <100$),
which are of practical interest to this work, it holds that:
\begin{equation}
CT\left(\left\Vert w\right\Vert \right)\sim2^{\left\Vert w\right\Vert }\label{eq:CTanyWsize}
\end{equation}
Corresponding expressions exist for random streams modeled as Markov
chains rather than i.i.d. processes~\cite{McConnel.2001}. The i.i.d.
assumption is retained for ease of exposition.

\section{The BitSurfing Adapter Model\label{sec:System-Model}}

The BitSurfing model proposes a new kind of network adapter that does
not generate new physical data packets when transmitting information.
Instead, it assigns meaning to symbols created by an external source
as described below. It is noted that the operation of the novel adapter
is transparent to the applications, which receive and request the
delivery of data as usual, i.e., without being aware of the nature
of the underlying adapter.

\begin{figure}[t]
\begin{centering}
\includegraphics[bb=0bp 0bp 820bp 600bp,width=1\columnwidth]{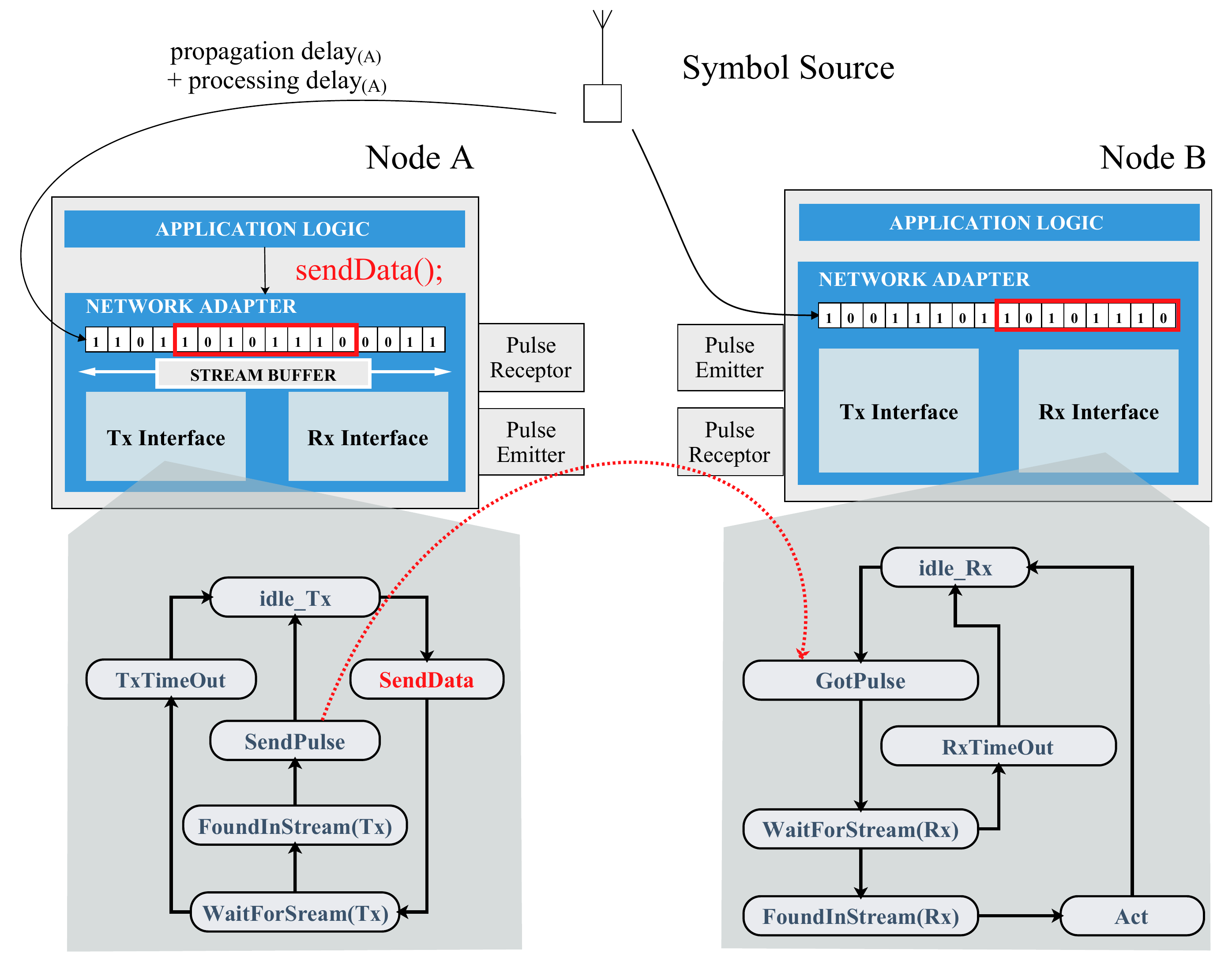}
\par\end{centering}
\caption{\label{fig:GrandPlan}Overview of the BitSurfing communication paradigm. }
\end{figure}

The proposed workflow is illustrated in Fig~\ref{fig:GrandPlan},
involving two nodes equipped with BitSurfing network adapters and
one symbol source. The source constantly produces new symbols and
each BitSurfing adapter reads and stores them in a finite-sized, FIFO
\emph{stream buffer}. When the application logic of Node A requests
the sending of data (i.e., a \emph{word}) to Node B, the adapter waits
until the word appears in its stream buffer. Then, it emits a single,
short pulse (i.e., a single bit), which acts as a notification to
Node B to scan its stream buffer for a valid word (i.e., contained
within a predefined, common \emph{codebook}). The first valid word
found is passed on to the application logic of Node B.

We proceed to make note of some important aspects of the described
workflow. Firstly, note that the stream buffers of two nodes may not
be identical at a given time moment. Propagation delays and symbol
processing time variations (for reading and storing symbols) may lead
to relatively shifted buffer states. Secondly, the adapter workflow
includes optional timeouts, to preclude that the adapter remains busy
beyond a certain desired duration. The time for such events can be
measured by counting symbol store events (in the buffer stream), rather
than with a regular clock. Finally while a sender can enqueue simultaneous
data send requests and treat them in a serial manner, a receiver can
only respond to a received pulse if it is in an idle state. Pulses
received by a busy receiving (Rx) interface are ignored.

Based on the described workflow, we make two important remarks:

$\blacksquare$~\emph{The BitSurfing workflow is event-driven.}

This statement reflects the fact that the workflow does not require
a hardware clock to time its operation. The adapter simply reacts
to three simple events, i.e., the arrival of a new symbol, a $\texttt{SendData()}$
request from the application, or the reception of a pulse. These attributes
make the BitSurfing architecture eligible for implementation as an
asynchronous chip. It is noted, however, that deriving specific hardware
designs is an open challenge.

$\blacksquare$~\emph{The symbol stream reception can vary across
the network, without impacting the BitSurfing performance.}

In order for two nodes to communicate successfully, their symbol buffer
contents may be relatively shifted, but should be otherwisely identical.
However, this condition needs only to hold \emph{per network link}.
This means that the symbol reception can vary across the network in
general. \emph{The only sufficient condition for BitSurfing is that
neighboring adapters derive the same (albeit shifted) bit stream}.

\section{Designing the BitSurfing Codebooks\label{sec:systemdesign}}

In this Section we study the characteristics of word sets (i.e., the
codebooks) that can be used for BitSurfing-based communications. The
driving design goals are perpetual operation and payload maximization.

\subsection{Perpetual operation}

\label{subsec:Perpetual-operation}

Perpetual operation is a major concern for nano-IoT networks. The
power budget of each nanonode should remain positive, meaning that
energy harvesting rate must at least match the energy expenditure
rate. A BitSurfing adapter expends energy to read and process the
symbol stream incoming from the source, and to send the occasional
pulse, as described in Section~\ref{sec:System-Model}. However,
experimental studies have shown that wireless reception and data processing
can be perpetually powered by the received carrier itself, without
the need for a battery or any other power supply~\cite{Tabesh.}.
Thus, we will assume that the BitSurfing adapter power consumption
stems only from the occasional pulse emissions. It is noted that the
pulses themselves could be potentially powered by the carrier as well.
Nonetheless, in absence of experimental data, we will assume the more
common case of the pulse emission system proposed by Jornet et al~\cite{Jornet.2012b},
which employs graphene nanoantennas for pulse creation and zinc-oxide
nanowires for energy harvesting.

Let $h$ denote the energy harvesting rate of a node, and let $\epsilon$
be the energy expended for transmitting a single pulse per word of
size $\left\Vert w\right\Vert $. The cover time is given by equation
(\ref{eq:CTanyWsize}), measured in bits, or $\nicefrac{2^{\left\Vert w\right\Vert }}{r}$
measured in $sec$, where $r$ is the bitrate of the symbol source
(in $bps$). Perpetual operation requires that $h$ is greater than
the power drain, i.e.:
\begin{equation}
h\ge\frac{\epsilon}{\frac{2^{\left\Vert w\right\Vert }}{r}}\Rightarrow\left\Vert w\right\Vert \ge\left\Vert w\right\Vert _{min},\,\left\Vert w\right\Vert _{min}=\left\lceil \log_{2}\frac{\epsilon\cdot r}{h}\right\rceil \label{ineq:perpetual}
\end{equation}

$\blacksquare$ \emph{For a source bitrate $r$ there exists a minimal
word size $\left\Vert w\right\Vert $ that yields perpetual operation.}

\begin{figure}[t]
\begin{centering}
\includegraphics[bb=0bp 0bp 1000bp 500bp,width=0.9\columnwidth]{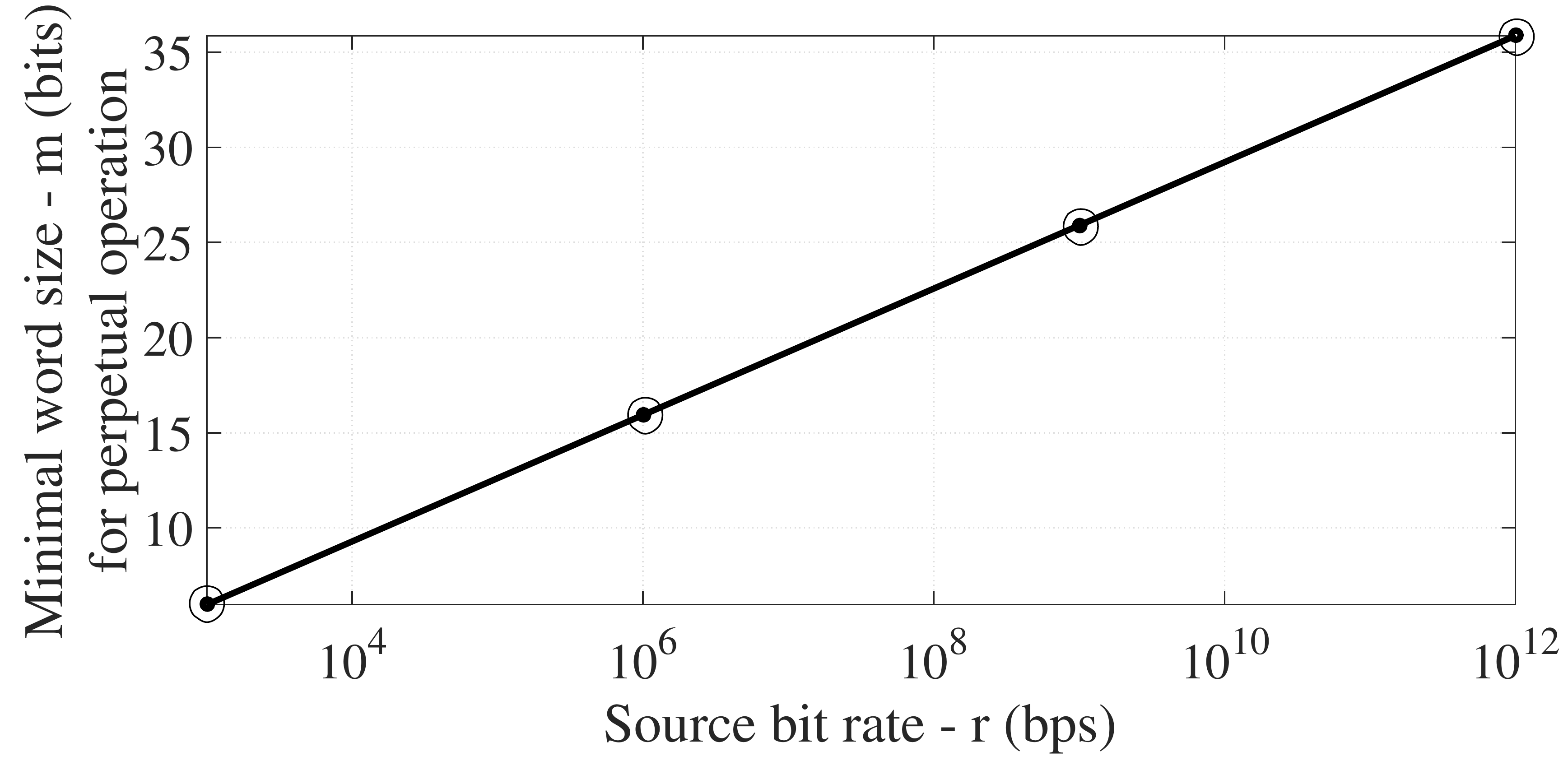}
\par\end{centering}
\caption{\label{fig:MinWordSizeForPerpet}The minimal word size for perpetual
operation under varying source bit rates. The $Kbps$, $Mbps$, $Gbps$
and $Tbps$ cases are highlighted.}
\end{figure}

Figure~\ref{fig:MinWordSizeForPerpet} shows the minimal word sizes
for perpetual operation, for a range of bit source rates, assuming
some common values for $\epsilon=1\,pJ$ and $h=16\,pJ/sec$~\cite{Jornet.2012b}.
Indicatively, the $r=1\,Kbps$, $1\,Mbps$, $1\,Gbps$ and $1\,Tbps$
cases correspond to $6$, $16$, $26$ and $36$-bit words.

\begin{figure}[t]
\begin{centering}
\includegraphics[bb=0bp 0bp 1000bp 500bp,width=0.9\columnwidth]{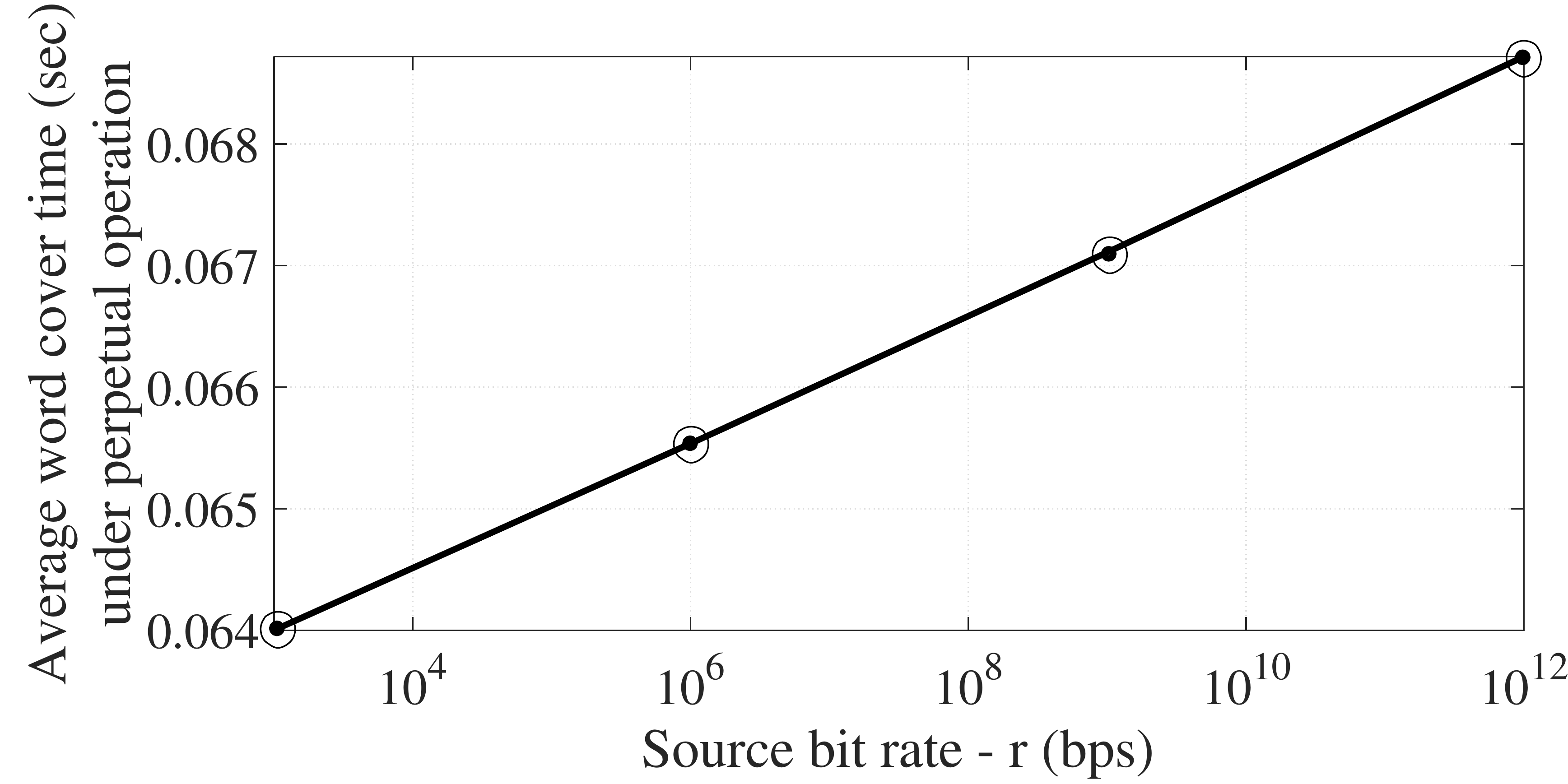}
\par\end{centering}
\caption{\label{fig:AcgCoverForPerpet}The average word covering time under
perpetual operation. The minimal word sizes of Fig.~\ref{fig:MinWordSizeForPerpet}
are considered for each source bit rate case. }
\end{figure}

It is interesting to note that the cover time (measured in $sec$)
corresponding to the minimal word sizes of relation (\ref{ineq:perpetual})
is approximately constant. This outcome follows from equation (\ref{eq:CTanyWsize}),
when $\left\Vert w\right\Vert =\left\Vert w\right\Vert _{min}$:
\begin{equation}
CT\left(\left\Vert w\right\Vert _{min}\right)=\frac{2^{\left\lceil \log_{2}\frac{\epsilon\cdot r}{h}\right\rceil }}{r}\sim\frac{\epsilon}{h}\label{eq:CTanyWsize-1}
\end{equation}
For instance, as shown in Fig.~\ref{fig:AcgCoverForPerpet}, the
expected word cover time is approximately $66\,msec$ for the studied
source rates of $1\,Kbps-1\,Tbps$, while $\nicefrac{\epsilon}{h}$
yields $1/16\,sec=62.5\,msec$.

Notice that the $62.5\,msec$ cover time value may constitute a worst-case
scenario for BitSurfing. These values are derived for ambient energy
harvesting~($\epsilon=1\,pJ$ and $h=16\,pJ/sec$~\cite{Jornet.2012b})
and not for the carrier-feed operation intended in this work~\cite{Tabesh.}.
While the $h$ value corresponding to the latter cannot be derived
without an actual implementation at nano-scale, it is expected that
it will be much higher than ambient energy harvesting. Carrier feed
is a type of wireless power transfer, which is not only more structured,
but can also be freely set to a conveniently high level. Higher $h$
values can reduce the outcome of eq. (\ref{eq:CTanyWsize-1}) even
to $\mu sec$ levels.

\subsection{Payload maximization\label{subsec:Payload-maximization}}

As described in Section~\ref{sec:System-Model}, searching the stream
buffer for valid words is the core operating principle of the BitSurfing
adapter. For reasons of hardware simplicity and search efficiency,
the codebook (i.e., the set of valid words) should support the self-synchronizing
property~\cite{varshavsky1990self}. This means that a search within
a binary string $S$ comprising any concatenation of codebook words
$S:\left\{ w_{1}+w_{2}+\ldots\right\} $, yields only the words $w_{1},\,w_{2},\ldots$,
and in the exact concatenation order. An approach for achieving this
property is described by the following remark:

$\blacksquare$~\emph{A codebook comprising words that: i) begin
with a prefix $p$, and ii) $p$ is not found anywhere else within
each word, is self-synchronizing.}

The prefix approach has additional advantages:

$\bullet$ The buffer search can only look for $p$ and then be triggered
to check if the remaining $\left\Vert w\right\Vert -\left\Vert p\right\Vert $
bits form a valid word.

$\bullet$ The search can occur at a fixed index position, since the
buffer contents move towards the FIFO direction. (Eventually, the
word will reach the fixed index position).

$\bullet$ Invalid words are easily discarded: once the $p$ bits
are found, they should not be encountered again for the remaining
$\left\Vert w\right\Vert -\left\Vert p\right\Vert $ bits. If they
are indeed encountered again, the adapter can safely discard and ignore
the previous occurrence of $p$ (forward detection of valid words).
\begin{algorithm}
\hspace*{\algorithmicindent}\textbf{Inputs}: A specific prefix, $p$ (e.g., '1101'); a word size $\left\Vert m\right\Vert$.\\
\hspace*{\algorithmicindent}\textbf{Output}: The codebook size, $max\_size$.

\begin{algorithmic}[1]

\STATE  \textbf{var} $max\_size\gets2^{\left\Vert m\right\Vert -\left\Vert p\right\Vert };$

\STATE  \textbf{for all} binary words $n$ so as $\left\{ \left\Vert n\right\Vert =\left\Vert m\right\Vert -\left\Vert p\right\Vert \right\} $

\STATE  \textbf{~~if} $\texttt{wrdfind}\left(\texttt{wrdcat}\left(p,\,n\right),\,p\right)>0$

\STATE  ~~~~$max\_size\gets max\_size-1;$

\STATE  \textbf{~~end if}

\STATE  \textbf{end for}

\rule[0.5ex]{0.9\columnwidth}{0.1pt}

{\scriptsize{}$\texttt{wrdcat}\left(a,\,b\right):$ concatenates words
$a$, $b$.}{\scriptsize \par}

{\scriptsize{}$\texttt{wrdfind}\left(a,\,b\right):$ returns the index
$\left(\ge0\right)$ of the }{\scriptsize{}\uline{last}}{\scriptsize{}
occurrence of word $b$ in word $a$, or $-1$ if not found.}{\scriptsize \par}

\end{algorithmic}

\caption{\label{alg:ALG1}Process for calculating the codebook size.}
\end{algorithm}
\begin{figure}[t]
\begin{centering}
\includegraphics[bb=50bp 0bp 1050bp 500bp,width=0.9\columnwidth]{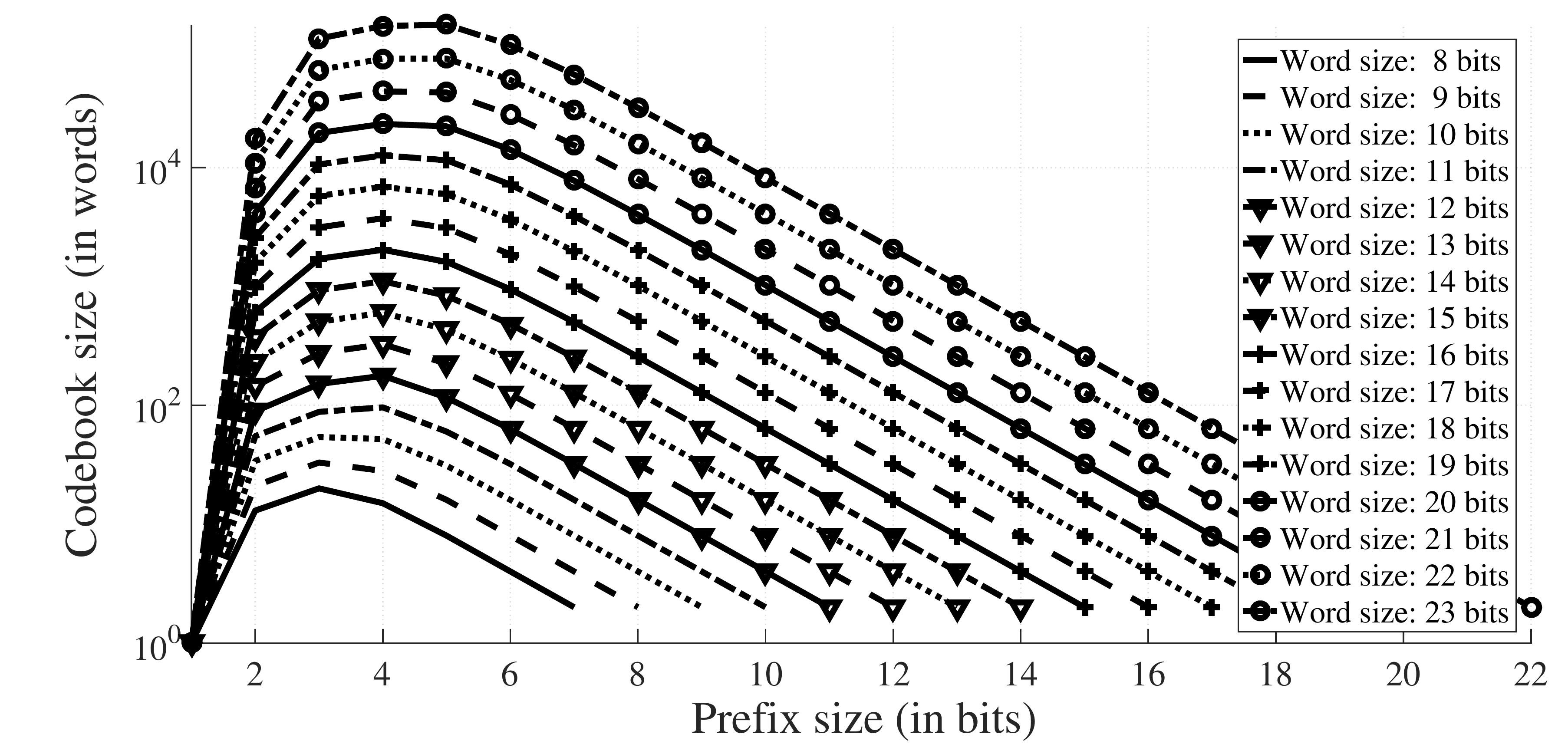}
\par\end{centering}
\caption{\label{fig:codebook_size_per_prefix}The codebook size (in number
of valid words) for varying word prefix bits. Several total word sizes
are presented.}
\end{figure}
\begin{figure}[t]
\begin{centering}
\includegraphics[bb=0bp 0bp 1000bp 500bp,width=0.9\columnwidth]{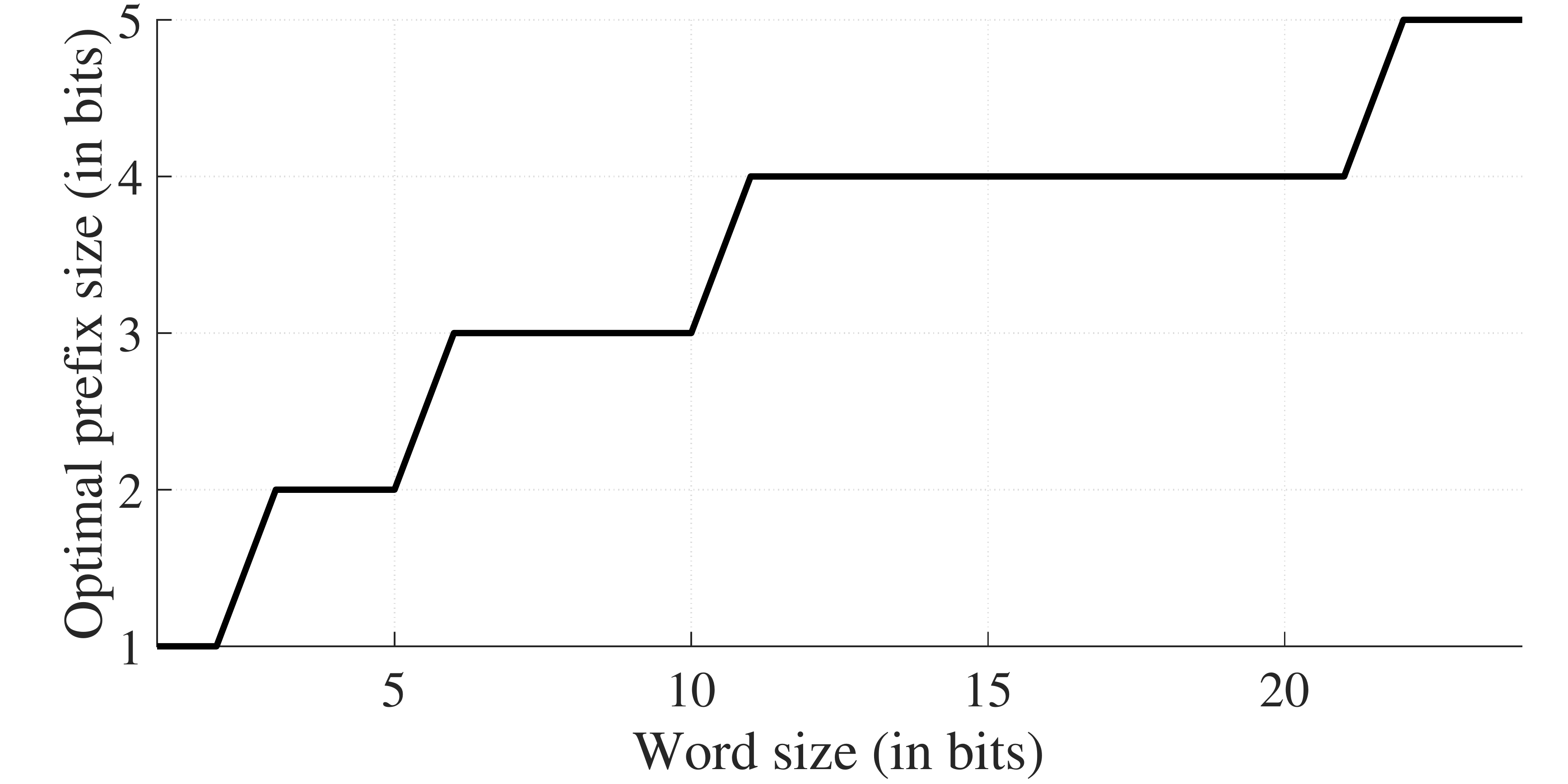}
\par\end{centering}
\caption{\label{fig:VErtCutOptimalPrefixes}Vertical cut of Fig.~\ref{fig:codebook_size_per_prefix}
showing the optimal prefix size (i.e., yielding the largest codebook)
per word size. }
\end{figure}

However, a disadvantage of the prefix approach is that it naturally
limits the number of valid words (and, hence, the payload per word),
making them less than the total $2^{\left\Vert w\right\Vert }$ possible
combinations. The process denoted as Algorithm~\ref{alg:ALG1} provides
a straightforward way for calculating the size of a valid codebook,
for a given prefix and a word size. Figure~\ref{fig:codebook_size_per_prefix}
illustrates the largest codebook sizes for several word and prefix
sizes. It is shown that for every word size there exists a prefix
size that offers the largest codebook. This is also illustrated in
Fig.~\ref{fig:VErtCutOptimalPrefixes}\textendash a vertical cut
of Fig.~\ref{fig:codebook_size_per_prefix}\textendash which summarizes
the optimal prefix sizes for given word sizes. For instance, words
with $11\le\left\Vert w\right\Vert \le21$ yield an optimal prefix
of size $4$.
\begin{figure}[t]
\begin{centering}
\includegraphics[bb=50bp 0bp 1300bp 500bp,width=0.9\columnwidth]{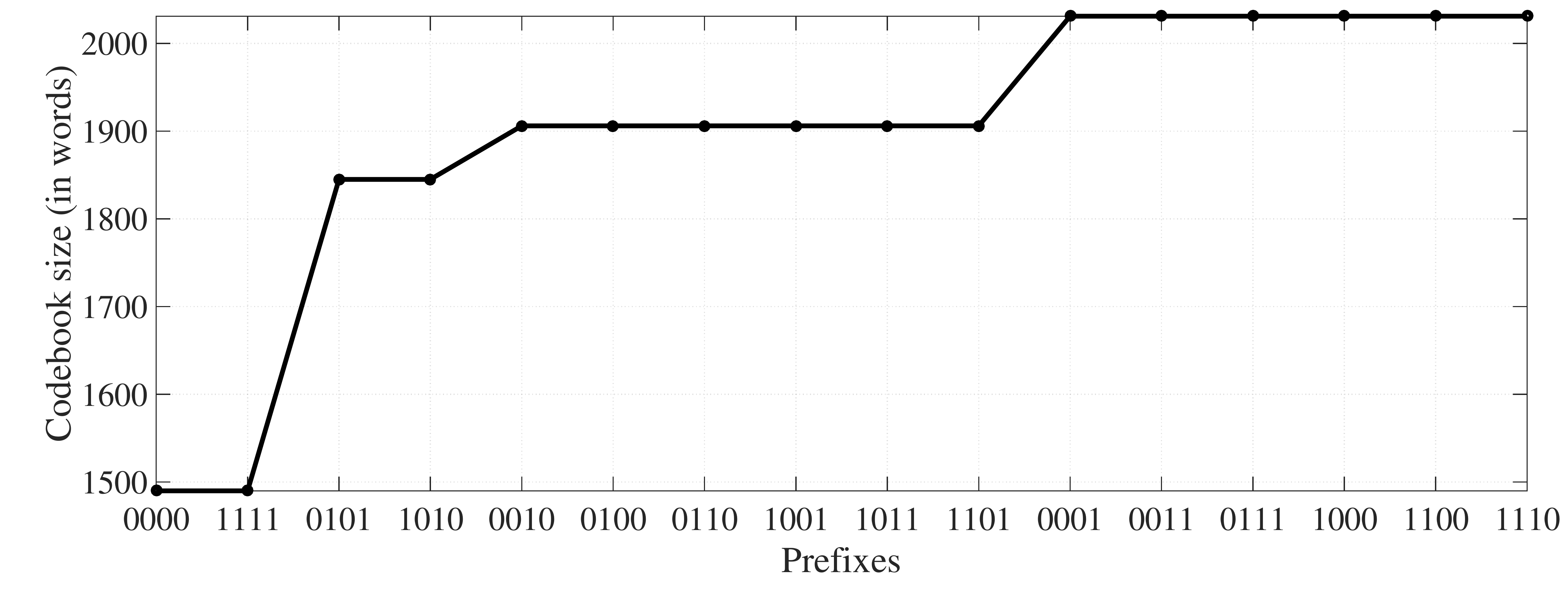}
\par\end{centering}
\caption{\label{fig:caseStudy16bitsPrefixes}Case study of 4-bit prefixes for
16-bit words. The six rightmost prefixes yield the largest codebooks. }
\end{figure}

Not all prefixes of equal size yield the same codebook size. Figure~\ref{fig:caseStudy16bitsPrefixes}
shows the outputs of Algorithm~\ref{alg:ALG1} for $16$-bit words
and all possible $4$-bit prefixes. The last six prefixes ($'0001'$
to $'1110'$) yield the largest codebook size of $\sim2100$ words.
\begin{figure}[t]
\begin{centering}
\includegraphics[bb=0bp 0bp 1000bp 500bp,width=0.9\columnwidth]{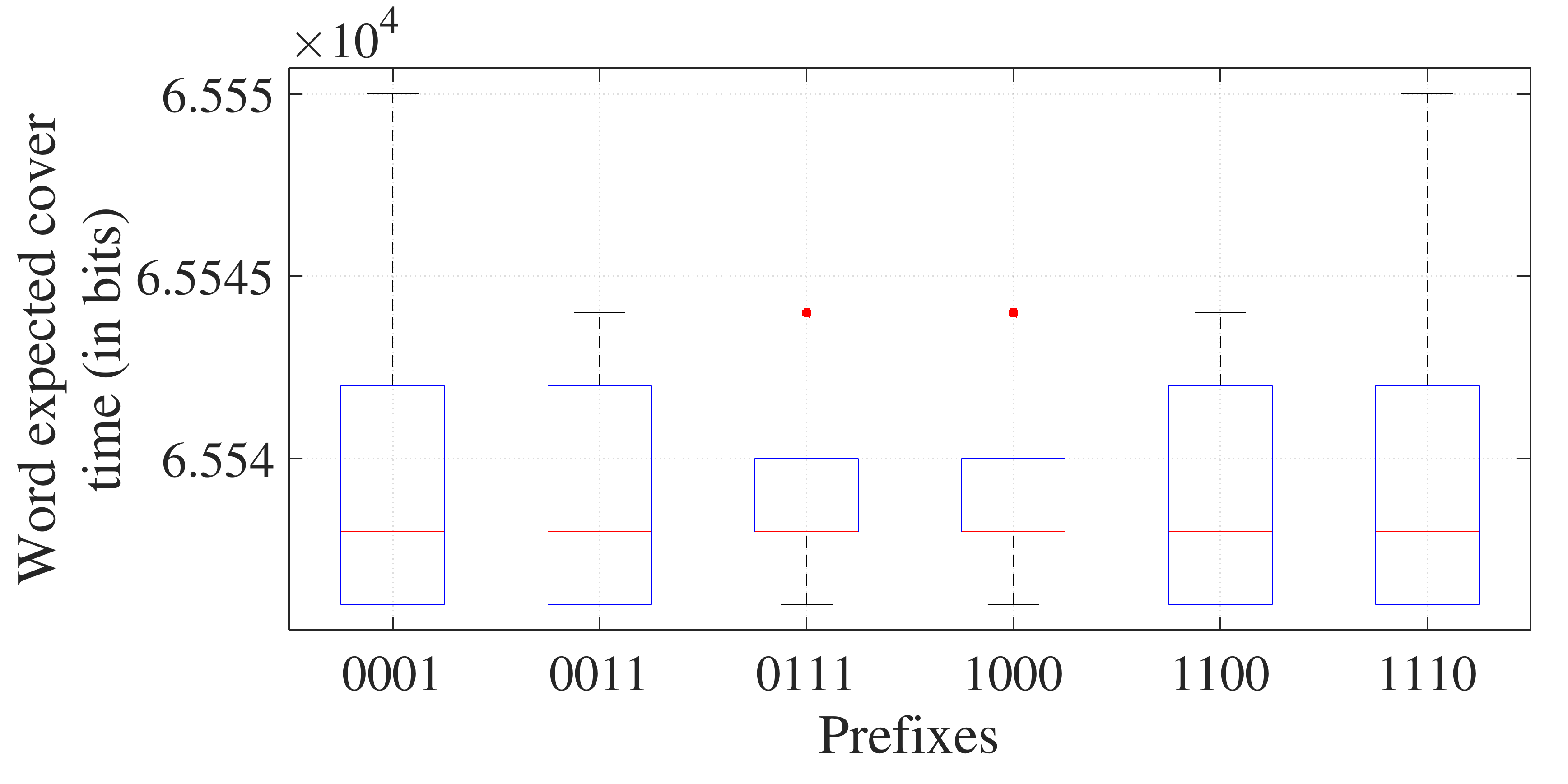}
\par\end{centering}
\caption{\label{fig:boxplots}Word cover time distribution for the codebooks
pertaining to the six best prefixes of Fig.~\ref{fig:caseStudy16bitsPrefixes}.
The '0111' and '1000' prefix cases offer the (marginally) smallest
cover time deviation.}
\end{figure}

One final consideration for choosing the best prefix (and the corresponding
codebook from Algorithm~\ref{alg:ALG1}) is the cover time distribution
of the words it contains. As shown by equation~(\ref{eq:CTw}), each
word generally yields a different cover time. Thus, in Fig.~\ref{fig:boxplots}:
i) we obtain the codebooks for each of the $'0001'$ to $'1110'$
prefixes of Fig.~\ref{fig:caseStudy16bitsPrefixes} using Algorithm~\ref{alg:ALG1},
and ii) calculate the expected cover time of each word therein. The
box-plots express the cover time distribution for each codebook/prefix.
The '0111' and '1000' prefixes exhibit the smallest cover time variance
(albeit a marginal one), and can both be chosen to produce codebooks
for the studied, $16$-bit word case.

Having described the process of designing the BitSurfing adapters
and codebooks, we proceed to evaluate a complete nano-IoT network
setup with BitSurfing adapter-equipped nodes.

\subsection{Security}

Security is considered as one of the main concerns in nano-IoT~\cite{Dressler.2012}.
Authorization and authentication is paramount for mission-critical
applications such as in medicine (e.g., in-body nano-IoT) and in mission-critical
industry (structural control of materials). However, extreme hardware
restrictions do not allow for a classic, cryptography-based approach.
In that sense, new approached to security are required.

Symbol generation outsourcing provides a degree of novelty that can
facilitate security. The node codebooks can be customized per application
instance, naturally containing the impact of potential hacking. Moreover,
hard-wired implementation at nano-scale means that the codebooks are
naturally protected against direct tampering (e.g., capturing and
reverse-engineering a nano-IoT node). Finally, the dependence on an
external power source naturally solves the problem of emergency shutting
down a nano-IoT network. This can be directly accomplished by removing
or powering off the source, rather than relying on some protocol mechanism
that could fail for a multitude of common causes.

\section{Evaluation \label{sec:sim}}

\begin{figure}[t]
\begin{centering}
\includegraphics[bb=0bp 0bp 400bp 350bp,clip,width=0.6\columnwidth]{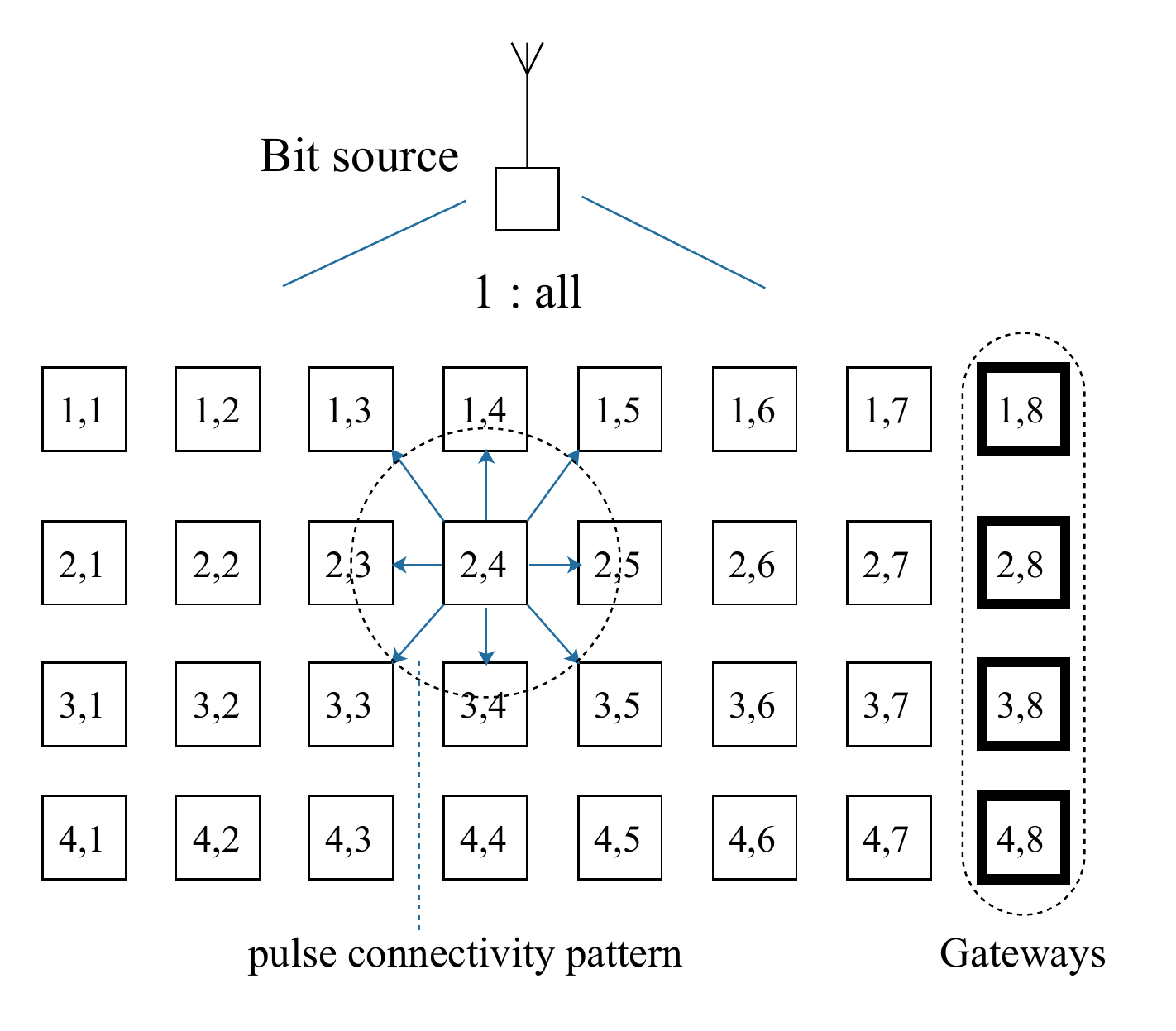}
\par\end{centering}
\caption{\label{fig:simsetup}The simulation setup, comprising one source and
$32$ nodes.}
\end{figure}
The following setup is simulated at bit-level in JAVA, using the AnyLogic
discrete event modeling platform~\cite{XJTechnologies.2013}. The
platform is based on Eclipse, and provides state-of-the-art facilities
for code visualization, general purpose optimization and automatic
statistical evaluation of models. A free version of the platform is
available for personal use~\cite{XJTechnologies.2013}. The runs
took place on commodity hardware (Intel i7 4770, 16GB DDR3). The simulation
files are freely available upon request.

The evaluation considers a multi-hop nano-IoT network with identical,
BitSurfing adapter-equipped nodes. The studied communication scenario
is applicable to HyperSurfaces, a novel class of planar materials
that can interact with impinging electromagnetic waves in a software-defined
manner~\cite{cacm}. They constitute a merge of nano-IoT and metamaterials~~\cite{Holloway.2012}.
Metamaterials comprise a two-dimensional pattern of a conductive material,
the\emph{ meta-atom}, repeated periodically over a dielectric substrate.
The form of the meta-atoms defines the electromagnetic response of
the surface, exemplary including the reflection of the impinging wave
at a custom angle (even at negative ones), full absorption, etc. The
\emph{HyperSurface} concept takes the metasurface concept one step
further: it allows the formation of custom meta-atoms over it. A nano-IoT
network embedded within the HyperSurface acts as the meta-atom sense
and control factor. It senses and sends useful attributes of impinging
waves to an external entity and receives back commands to ``draw''
meta-atoms by altering the local conductivity of the HyperSurface
accordingly. For ease of exposition the evaluation focuses on the
sensing direction, noting that the actuation direction is similar.
With ease-of-manufacturing in mind, the source rate is set to $r=1\,Mbps$.

\textbf{Setup}. We assume the system model of Fig.~\ref{fig:GrandPlan}.
Notice that, in accordance with Fig.~\ref{fig:MinWordSizeForPerpet},
the chosen rate $r=1\,Mbps$ corresponds to $16$-bit words in order
to achieve perpetual operation. Moreover, based on Fig.~\ref{fig:caseStudy16bitsPrefixes}
and \ref{fig:boxplots}, we adopt the '1000' prefix which corresponds
to a codebook of size $2100$, i.e., $11$ bits for payload per word
(since $\log_{2}\left(2100\right)\approx11$). We will use these bits
as follows: identifier of a sender node ($5$-bits), identifier of
a recipient node ($5$-bits), and measurement data ($1$-bit) exemplary
expressing whether the sensed current within a meta-atom surpasses
a threshold. $5$-bit identifiers can uniquely express $2^{5}=32$
nodes, which explains the topology size of Fig.~\ref{fig:GrandPlan}.
(We note again that this limitation is due to the runtime consideration
only). The node identifiers are considered hard-coded and well-arranged,
as shown in the Figure. The four nodes placed at the right-most locations
act as gateways to the external world. The pulse connectivity range
shown is intended to increase the hops required for reaching a gateway,
making successful delivery more challenging. As a general note, smaller
pulse range also translates to less energy per pulse emission.

Finally, the steam buffer size is set to $30$ bits, i.e., enough
to accommodate $16$-bits words and up to $14$-bit buffer shifts
due to variable propagation and processing delay. Each node has a
random such delay, picked uniformly. Timeouts are set to $10^{6}$
bits (i.e., 1 $sec$).

\textbf{Application logic}. When requested by the simulation, a node
crafts a data packet (i.e., a word), comprising his identifier, the
identifier of an immediate recipient node to its right picked at random,
and a random measurement data. For instance, in Fig.~\ref{fig:simsetup},
node $\left(2,4\right)$ picks any of the $(1,5)$, $(2,5)$, $(3,5)$
nodes at random as the immediate recipient. The packet is forwarded
to the BitSurfing adapter, which starts waiting for it to appear in
the stream buffer. If the adapter is busy, the packet is enqueued
until the adapter becomes idle. When receiving a packet (by getting
a pulse and retrieving the word from the stream buffer), a node first
checks if itself is the immediate recipient. If not, the packet is
ignored. If yes, it rewrites the immediate recipient list with one
of its own right-hand neighbors. Gateways act as packet consumers
only and do not create new ones.

\textbf{Run configuration}. We are interested in evaluating the BitSurfing-based
communication in terms of successful packet delivery ratio (i.e.,
reaching the gateways) under various congestion levels, while logging
transmission times owed to stream cover times and packet queuing delays.
To this end, a single run comprises $100$ successive packet creation
phases. At the start of each phase, a number of packets are created
simultaneously by randomly selected nodes. Naturally, more packets
created simultaneously result into a more congested network. Once
all packets have been delivered (or lost), a new creation phase begins.
The following results refer to all $100$ phases, to ensure high confidence
in the logged values.
\begin{table}
\caption{\label{tab:Packet-delivery-rate}Packet delivery rate versus network
congestion level (network packets created simultaneously).}
\centering{}%
\begin{tabular}{|c|c|}
\hline
Congestion & Packet delivery rate\tabularnewline
\hline
\hline
$1$ & $100\%$\tabularnewline
\hline
$10$ & $100\%$\tabularnewline
\hline
$20$ & $99.62\%$\tabularnewline
\hline
$40$ & $99.51\%$\tabularnewline
\hline
$60$ & $99.51\%$\tabularnewline
\hline
$80$ & $99.37\%$\tabularnewline
\hline
$100$ & $99.36\%$\tabularnewline
\hline
\end{tabular}
\end{table}

\begin{figure}[t]
\begin{centering}
\includegraphics[width=0.9\columnwidth]{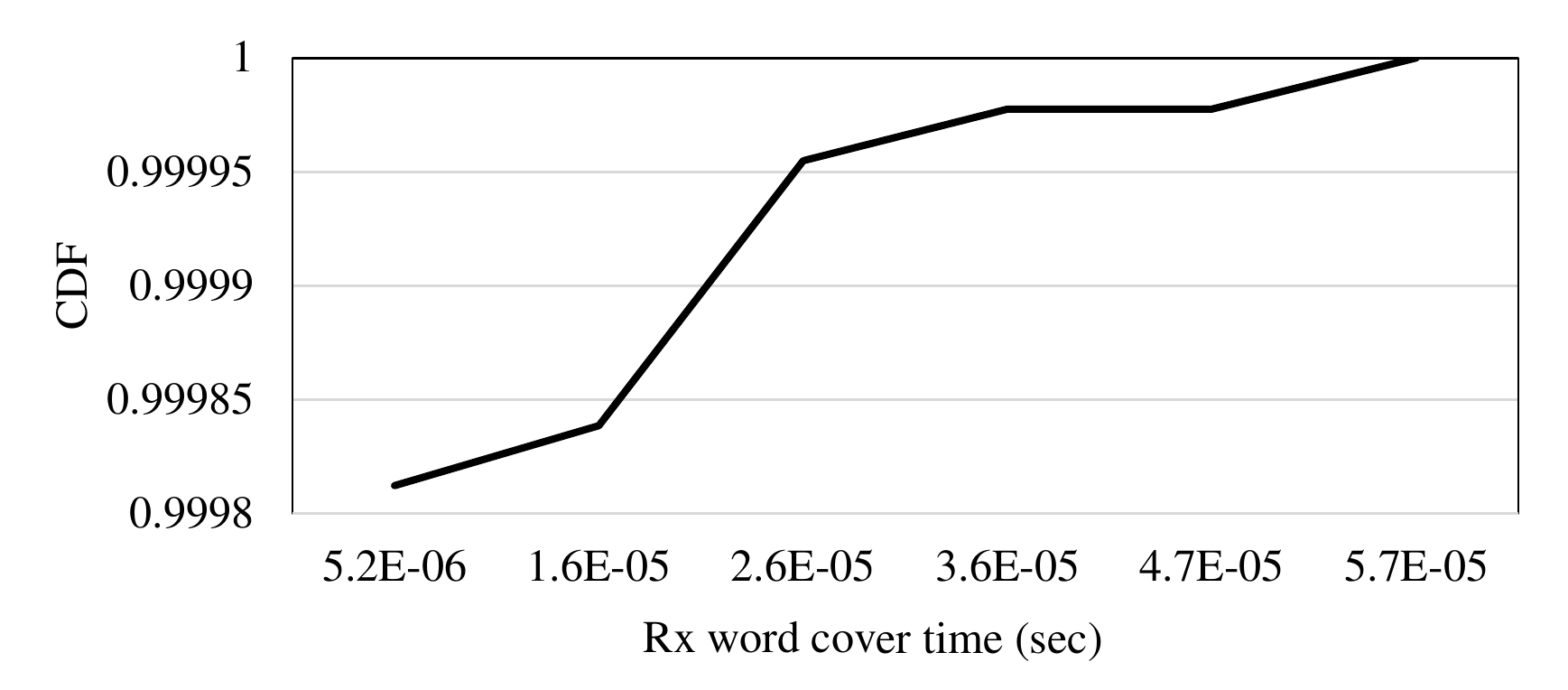}
\par\end{centering}
\caption{\label{fig:rxtimesCDF}CDF of the Rx word cover times (Congestion
level: 100)}
\end{figure}

\textbf{Results}. Table~\ref{tab:Packet-delivery-rate} presents
the delivery rates over all packet in each run, for different congestion
levels. It is shown that BitSurfing offers an almost perfect delivery
rate. The very few packet losses are owed to pulse collisions, i.e.,
when two pulses corresponding to different packets reach the same
immediate recipient node at nearly the same time. Such collisions
are highly unlikely to occur, given that pulses have $psec$ duration~\cite{Jornet.2012b},
while the Rx stream processing time is also trivial ($\mu sec$),
as shown later in Fig.~\ref{fig:rxtimesCDF}. This means that \emph{BitSurfing
adapters can operate without a medium access control mechanism, simplifying
their hardware implementation.}
\begin{figure}[t]
\begin{centering}
\includegraphics[width=0.9\columnwidth]{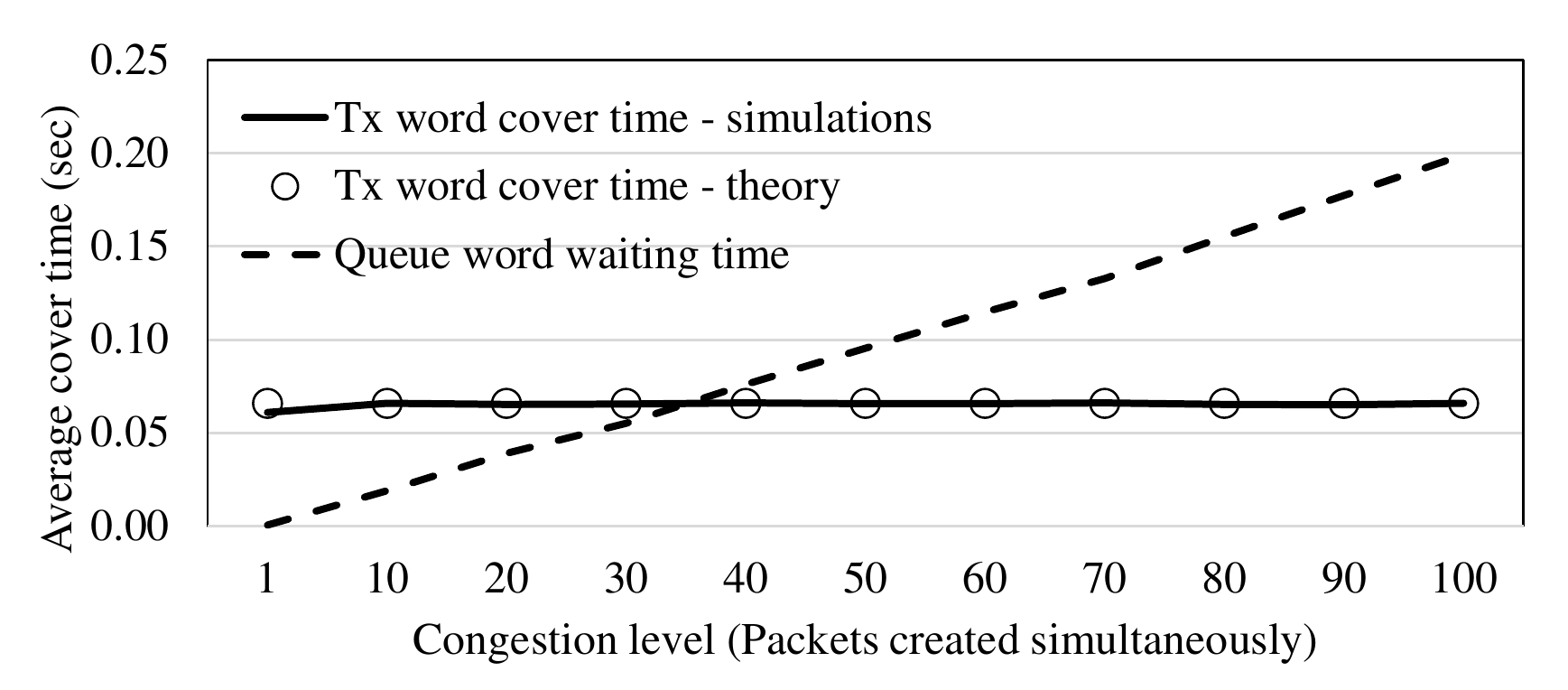}
\par\end{centering}
\caption{\label{fig:avgWTvsCong}Average time to traverse a network hop due
to Tx word cover time and word queuing delay, under varying network
congestion levels.}
\end{figure}

We proceed to study the transmission times in Fig.~\ref{fig:avgWTvsCong}.
The Figure illustrates the average time a word needs to cross a hop
between two nodes, owed to (Tx) stream cover time and queuing time.
Notably, \emph{the cover time is invariant to network congestion},
as expected by equations (\ref{eq:CTw}) and (\ref{eq:CTanyWsize}).
Moreover, the simulation are in agreement with the theoretical expectation
(i.e. $66\,msec$) expressed by these equations. Thus, from a network
congestion level and on, the queuing time becomes the dominant factor
in the propagation delay.
\begin{figure}[t]
\begin{centering}
\includegraphics[width=1\columnwidth]{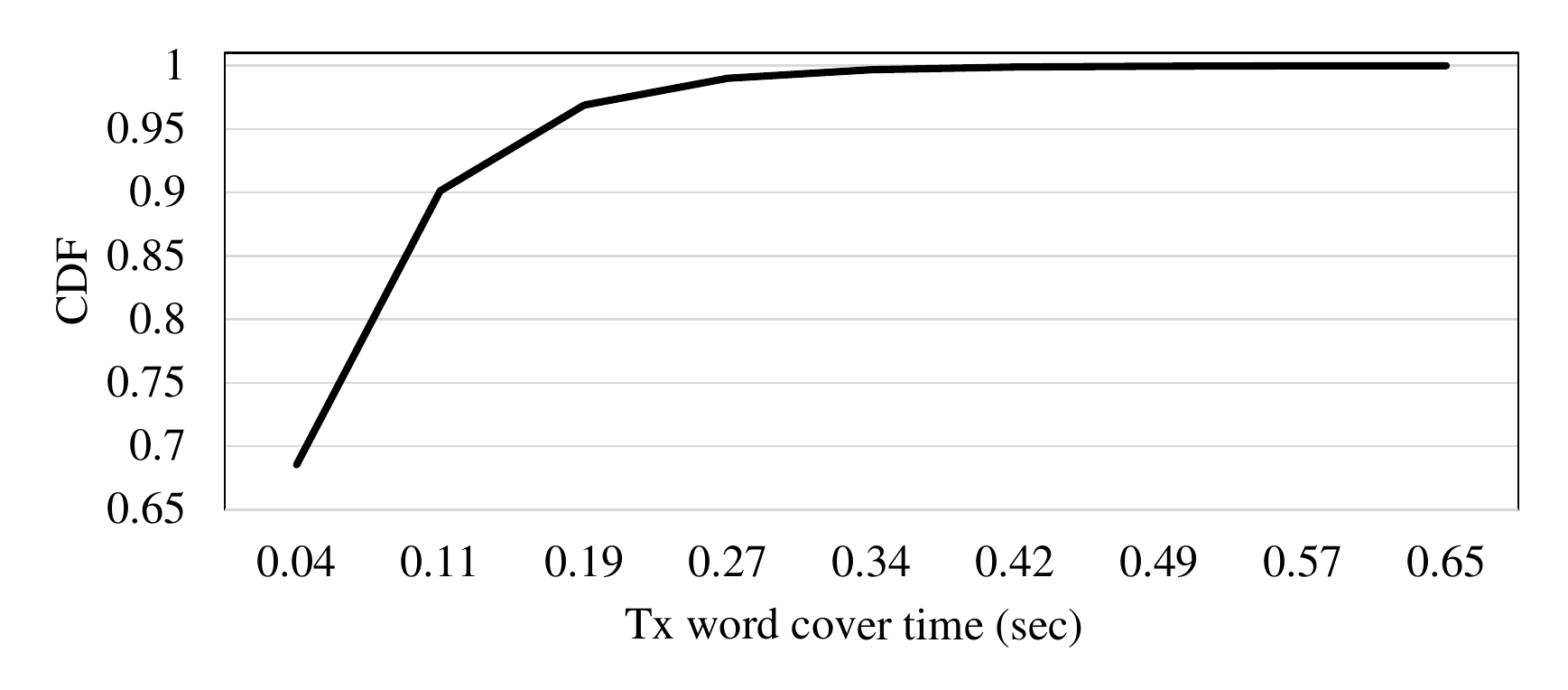}
\par\end{centering}
\caption{\label{fig:TxTimesCDF}CDF of the Tx word cover times (Congestion
level: 100).}
\end{figure}

Figure~\ref{fig:TxTimesCDF} proceeds to detail the CDF of the Tx
cover times. It can be seen that there is a $80\%$ probability that
the cover time will be less than the average value of $66\,msec$.
Moreover, the cover time is almost certainly less than five times
the average, i.e., $330\,msec$.
Finally, it is worth noting that the Rx cover times are very low,
as shown in the CDF of Fig.~\ref{fig:rxtimesCDF}. When a receiving
node gets an incoming pulse, the intended word is either already within
the stream buffer or about to enter it. Thus, the Rx interface remains
busy for very small time intervals ($\mu sec$) before returning to
the idle state. Thus, the collision probability becomes trivial, even
for wireless networking standards, as shown in Table~\ref{tab:Packet-delivery-rate}.

\section{Discussion and Research Directions\label{sec:directions}}

BitSurfing communications were shown to exhibit some interesting benefits,
summarized as follows:

$\bullet$ Simplified transmitter hardware: No packet transmission
circuitry and potential for clock-less implementation.

$\bullet$ Simplified power supply: The BitSurfing adapters are intended
to be fully carrier-fed and, thus, be completely battery-less and
perpetually powered. This prospect is strongly supported by related
implementations~~\cite{Tabesh.}.

$\bullet$ Collision-less communication: In BitSurfing, a data packet
exchange corresponds to one ultra-short pulse emission, regardless
of the data packet size. Thus, collisions are practically inexistent,
even at fully congested multi-hop networks.

However, BitSurfing requires a different hardware/software development
style. Instead of approaching the hardware, the protocols and the
application logic in a disjoint manner\textendash i.e., the current
common practice\textendash BitSurfing dictates a close co-design process.
The maximum number of nodes, the data latency requirements, the intended
data format, all affect each other and define strict conditions that
should be upheld by the BitSurfing hardware and protocols. Nonetheless,
the enforced co-design comes with a clear workflow that was described
in Section\ \ref{sec:systemdesign}. This workflow readily provides
the available BitSurfing parameterization corresponding to any requirements.
There, it can actually guide and simplify the implementation process.

The co-design workflow and the hardware simplification benefits of
BitSurfing constitutes it a promising approach for nano-IoT. Nanonodes
are extremely restricted in terms of dimensions, which naturally calls
for careful software/hardware co-design for optimal usage of the available
space. The potential for operation without battery, transmission circuitry
and MAC protocol facilitates miniaturization further. A consideration,
however, is that the application scenario must allow for the presence
of the external source that generates symbols and feeds the nanonodes
with energy.

Finally, regarding the relation to existing studies, the BitSurfing
model is novel, to the best of the authors' knowledge. A different
concept that exhibits some conceptual similarity to BitSurfing is
the \emph{communication through silence} (CtS)~\cite{zhu2005challenges}.
CtS is ad-hoc, without external sources. To transmit a packet, a node
emits a pulse to its neighbor. The recipient then starts counting
from $0$ in unary steps. The sender emits another, carefully synchronized
pulse that notifies the recipient to stop counting and interpret the
reached number as data. Collisions are very often: if another sender
emits a pulse while the recipient is counting (a time-consuming task),
it is still interpreted as a stop-counting signal. Moreover, perfect
clocks are needed: if a pulse slips even by a single timeslot (e.g.,
$\pm1$ clock tick), the perceived data will be erroneous. Thus, CtS
does not exhibit the aforementioned benefits of BitSurfing in the
context of nano-IoT.

\subsection{Research Directions}

BitSurfing opens several interesting research directions, listed per
adapter component:

\textbf{Symbol source}. The present study assumed an i.i.d. generated
binary stream. Future extensions can study the cover time in existing
and widely-used symbol sources, such as WiFi, Cellular and DVB/T~\cite{ho2017game}.

Moreover, extensions can study symbol sources specially designed for
BitSurfing, rather than opportunistic ones. For instance, such a source
can be restricted to generate and broadcast only valid words, rather
than random bits. The schedule of such broadcasts and its adaptivity
to the traffic pattern of the nanonodes is another open direction.
Studies on broadcast scheduling can constitute the starting basis
for this direction~\cite{Liaskos.2012}.

\textbf{Adapter codebook}. The preceding analysis and simulations
considered codebooks with equi-sized words. Novel codebooks with variable
size words can be developed, that match the traffic characteristics
of the nano-IoT network~\cite{lam1992self}. For instance, smaller
words (with smaller cover time) can be used to represent acknowledgment
messages. This direction can be generalized as protocol/codebook co-design.

Additionally, codebooks can be designed to offer robustness against
stream symbol reception errors. For instance, the lexicographical
distance between words can be maximized~\cite{hu2011improved}, to
limit the probability that an erroneously received word will be treated
as another valid one. Model checking techniques can study the effects
of such events~\cite{volk2018fast}, subsequently proposing protocol
revisions as needed.

\textbf{Adapter hardware}. As discussed in Section~\ref{subsec:Perpetual-operation},
related studies provide strong evidence that the stream processing
part of the BitSurfing adapters can be perpetually powered, without
batteries~\cite{Tabesh.}. Hardware-oriented studies are required
to prove that the 1-bit pulse emissions can also be perpetually powered.
This is expected to depend on the pulse emission hardware, the pulse
propagation model~(e.g., \cite{wirdatmadja2018light}), and the required
pulse range. Note that multihop networks, like the one studied in
Section~\ref{sec:sim}, require very short-range pulses.

Depending on the available power budget, hardware implementation can
study ``colored'' pulses and multiple Rx and Tx interfaces per BitSurfing
adapter. New $\texttt{sendData()}$ requests from the application
layer can then be forwarded to the Tx interface with the fewest enqueued
requests, cutting down the word transmission times.

\textbf{Porting well-known protocols}. A promising point of the BitSurfing
paradigm is that it may remove constraints in porting well-known protocols
of wireless sensor networks to the nano-scale~\cite{sarangapani2017wireless}.
The perpetual and the MAC-less operation may allow for exchanging
any number of packets, enabling the adaptation of common addressing
and routing protocols (such as AODV) for nano-IoT. In the case of
codebooks comprising small words (e.g., the 16-bit case simulated
in Section~\ref{sec:sim}), data sessioning can undertake the task
of transparently breaking down long messages as required~\cite{abouzeid2003comprehensive}.

\section{Conclusion\label{sec:Conclusion}}

The present work proposed a novel nano-IoT network adapter named BitSurfer.
The novel adapter decouples the symbol generation from the communication
process. BitSurfing nanonodes rely on an external generator which
creates a continuous stream of symbols. The nanonodes read the generated
symbols awaiting the appearance of their intended messages, and then
using 1-bit, low-energy pulses to notify each other. The operation
of BitSurfing adapters is completely transparent to applications.
Moreover, BitSurfing offers significantly simplified nanonode transceiver
hardware, perpetual\textendash and potentially completely battery-less\textendash operation.
Furthermore, the novel adapters can operate without medium access
control, due to the very low probability of pulse collisions. The
multi-hop network with BitSurfing-enabled nanonodes was simulated,
exhibiting nearly perfect packet delivery rates and low delivery times.

\section*{Acknowledgment}

This work was funded by the European Union via the Horizon 2020: Future
Emerging Topics call (FETOPEN), grant EU736876, project VISORSURF
(http://www.visorsurf.eu).

{\small{}\bibliographystyle{IEEEtran}

\begin{thebibliography}{10}
\providecommand{\url}[1]{#1}
\csname url@samestyle\endcsname
\providecommand{\newblock}{\relax}
\providecommand{\bibinfo}[2]{#2}
\providecommand{\BIBentrySTDinterwordspacing}{\spaceskip=0pt\relax}
\providecommand{\BIBentryALTinterwordstretchfactor}{4}
\providecommand{\BIBentryALTinterwordspacing}{\spaceskip=\fontdimen2\font plus
\BIBentryALTinterwordstretchfactor\fontdimen3\font minus
  \fontdimen4\font\relax}
\providecommand{\BIBforeignlanguage}[2]{{%
\expandafter\ifx\csname l@#1\endcsname\relax
\typeout{** WARNING: IEEEtran.bst: No hyphenation pattern has been}%
\typeout{** loaded for the language `#1'. Using the pattern for}%
\typeout{** the default language instead.}%
\else
\language=\csname l@#1\endcsname
\fi
#2}}
\providecommand{\BIBdecl}{\relax}
\BIBdecl

\bibitem{cacm}
C.~Liaskos, A.~Tsioliaridou, A.~Pitsillides, S.~Ioannidis, and I.~F. Akyildiz,
  ``Using any surface to realize a new paradigm for wireless communications,''
  \emph{Communications of the ACM (to appear)}, 2018.

\bibitem{Jornet.2012b}
J.~Jornet and I.~Akyildiz, ``{Joint Energy Harvesting and Communication
  Analysis for Perpetual Wireless Nanosensor Networks in the Terahertz Band},''
  \emph{{IEEE Trans. on Nanotech.}}, vol.~11, no.~3, pp. 570--580, 2012.

\bibitem{limitedSpace}
F.-L.~A. Lau \emph{et~al.}, ``Computational requirements for nano-machines:
  There is limited space at the bottom,'' in \emph{NANOCOM'17}, pp. 1--6.

\bibitem{beerel2011proteus}
P.~A. Beerel, G.~D. Dimou, and A.~M. Lines, ``Proteus: An asic flow for ghz
  asynchronous designs,'' \emph{IEEE Design \& Test of Computers}, vol.~28,
  no.~5, pp. 36--51, 2011.

\bibitem{Tabesh.}
M.~Tabesh, M.~Rangwala, A.~M. Niknejad, and A.~Arbabian, ``A power-harvesting
  pad-less mm-sized 24/60ghz passive radio with on-chip antennas,'' in
  \emph{IEEE Symposium on VLSI Circuits}, 2014.

\bibitem{McConnel.2001}
T.~McConnel, ``The expected time to find a string in a random binary
  sequence,'' \emph{[Online] http://barnyard.syr.edu/cover.pdf}, 2001.

\bibitem{McConnel.1994}
------, ``cover.c: Compute the expected time to obtain ("cover") a given string
  of zeros and ones by a sequence of iid bernoullis.'' \emph{[Online]
  http://barnyard.syr.edu/quickies/cover.c}, 1994.

\bibitem{varshavsky1990self}
V.~I. Varshavsky, ``Self-synchronizing codes,'' in \emph{Self-Timed Control of
  Concurrent Proc.}\hskip 1em plus 0.5em minus 0.4em\relax Springer, 1990, pp.
  43--63.

\bibitem{Dressler.2012}
F.~Dressler and F.~Kargl, ``{Towards security in nano-communication},''
  \emph{{Nano Communication Networks}}, vol.~3, no.~3, pp. 151--160, 2012.

\bibitem{XJTechnologies.2013}
\BIBentryALTinterwordspacing
{XJ~Technologies}, \emph{{The AnyLogic Simulator}}, 2018. [Online]. Available:
  \url{http://www.anylogic.com}
\BIBentrySTDinterwordspacing

\bibitem{Holloway.2012}
C.~L. Holloway \emph{et~al.}, ``{An Overview of the Theory and Applications of
  Metasurfaces},'' \emph{{IEEE Ant. and Propag.}}, vol.~54, no.~2, pp. 10--35,
  2012.

\bibitem{zhu2005challenges}
Y.~Zhu and R.~Sivakumar, ``Challenges: communication through silence in
  wireless sensor networks,'' in \emph{The 11th annual international conference
  on Mobile computing and networking}.\hskip 1em plus 0.5em minus 0.4em\relax
  ACM, 2005, pp. 140--147.

\bibitem{ho2017game}
D.~Ho, G.~S. Park, and H.~Song, ``Game-theoretic scalable offloading for video
  streaming services over lte and wifi networks,'' \emph{IEEE Trans. on Mobile
  Computing}, 2017.

\bibitem{Liaskos.2012}
C.~Liaskos and G.~Papadimitriou, ``{G}eneralizing the {S}quare {R}oot {R}ule
  for {O}ptimal {P}eriodic {S}cheduling in {P}ush-based {W}ireless
  {E}nvironments,'' \emph{IEEE Trans. on Computers}, vol.~62, no.~5, pp.
  1044--1050, 2012.

\bibitem{lam1992self}
W.-M. Lam and A.~Reibman, ``Self-synchronizing variable-length codes for image
  transmission,'' in \emph{IEEE ICASSP'92}.

\bibitem{hu2011improved}
J.~Hu, H.~Zhang, Q.~Gao, and H.~Huang, ``An improved lexicographical sort
  algorithm of copy-move forgery detection,'' in \emph{IEEE ICNDC'11}.

\bibitem{volk2018fast}
M.~Volk, S.~Junges, and J.-P. Katoen, ``Fast dynamic fault tree analysis by
  model checking techniques,'' \emph{IEEE Trans. on Industrial Informatics},
  vol.~14, no.~1, pp. 370--379, 2018.

\bibitem{wirdatmadja2018light}
S.~Wirdatmadja \emph{et~al.}, ``Light propagation analysis in nervous tissue
  for wireless optogenetic nanonetworks,'' in \emph{Optogenetics and Optical
  Manipulation 2018}.

\bibitem{sarangapani2017wireless}
J.~Sarangapani, \emph{Wireless ad hoc and sensor networks}.\hskip 1em plus
  0.5em minus 0.4em\relax CRC Press, 2017.

\bibitem{abouzeid2003comprehensive}
A.~A. Abouzeid, S.~Roy, and M.~Azizoglu, ``Comprehensive performance analysis
  of a tcp session over a wireless fading link with queueing,'' \emph{IEEE
  Trans. on Wireless Comm.}, vol.~2, no.~2, pp. 344--356, 2003.

\end{thebibliography}

}{\small \par}
\end{document}